\documentclass[aps,prl,twocolumn,showpacs,groupedaddress,floatfix,superscriptaddress]{revtex4}
\usepackage{graphicx}
\usepackage[dvips]{color}
\usepackage{dcolumn}
\usepackage{bm}
\usepackage{amssymb}
\usepackage{amsmath}
\usepackage{multirow}
\usepackage{verbatim}
\usepackage{array}

\def\met{\mbox{${\hbox{$E$\kern-0.6em\lower-.1ex\hbox{/}}}_T$}}

\begin{document}
  
  \widetext
\hspace{5.2in} \mbox{FERMILAB-PUB-12-485-E}
  
  \title{\boldmath Limits on anomalous trilinear gauge boson couplings from 
    $WW$, $WZ$ and $W\gamma$ production in $p\bar{p}$ collisions at $\sqrt{s}=1.96$~TeV}

\affiliation{LAFEX, Centro Brasileiro de Pesquisas F\'{i}sicas, Rio de Janeiro, Brazil}
\affiliation{Universidade do Estado do Rio de Janeiro, Rio de Janeiro, Brazil}
\affiliation{Universidade Federal do ABC, Santo Andr\'e, Brazil}
\affiliation{University of Science and Technology of China, Hefei, People's Republic of China}
\affiliation{Universidad de los Andes, Bogot\'a, Colombia}
\affiliation{Charles University, Faculty of Mathematics and Physics, Center for Particle Physics, Prague, Czech Republic}
\affiliation{Czech Technical University in Prague, Prague, Czech Republic}
\affiliation{Center for Particle Physics, Institute of Physics, Academy of Sciences of the Czech Republic, Prague, Czech Republic}
\affiliation{Universidad San Francisco de Quito, Quito, Ecuador}
\affiliation{LPC, Universit\'e Blaise Pascal, CNRS/IN2P3, Clermont, France}
\affiliation{LPSC, Universit\'e Joseph Fourier Grenoble 1, CNRS/IN2P3, Institut National Polytechnique de Grenoble, Grenoble, France}
\affiliation{CPPM, Aix-Marseille Universit\'e, CNRS/IN2P3, Marseille, France}
\affiliation{LAL, Universit\'e Paris-Sud, CNRS/IN2P3, Orsay, France}
\affiliation{LPNHE, Universit\'es Paris VI and VII, CNRS/IN2P3, Paris, France}
\affiliation{CEA, Irfu, SPP, Saclay, France}
\affiliation{IPHC, Universit\'e de Strasbourg, CNRS/IN2P3, Strasbourg, France}
\affiliation{IPNL, Universit\'e Lyon 1, CNRS/IN2P3, Villeurbanne, France and Universit\'e de Lyon, Lyon, France}
\affiliation{III. Physikalisches Institut A, RWTH Aachen University, Aachen, Germany}
\affiliation{Physikalisches Institut, Universit\"at Freiburg, Freiburg, Germany}
\affiliation{II. Physikalisches Institut, Georg-August-Universit\"at G\"ottingen, G\"ottingen, Germany}
\affiliation{Institut f\"ur Physik, Universit\"at Mainz, Mainz, Germany}
\affiliation{Ludwig-Maximilians-Universit\"at M\"unchen, M\"unchen, Germany}
\affiliation{Fachbereich Physik, Bergische Universit\"at Wuppertal, Wuppertal, Germany}
\affiliation{Panjab University, Chandigarh, India}
\affiliation{Delhi University, Delhi, India}
\affiliation{Tata Institute of Fundamental Research, Mumbai, India}
\affiliation{University College Dublin, Dublin, Ireland}
\affiliation{Korea Detector Laboratory, Korea University, Seoul, Korea}
\affiliation{CINVESTAV, Mexico City, Mexico}
\affiliation{Nikhef, Science Park, Amsterdam, the Netherlands}
\affiliation{Radboud University Nijmegen, Nijmegen, the Netherlands}
\affiliation{Joint Institute for Nuclear Research, Dubna, Russia}
\affiliation{Institute for Theoretical and Experimental Physics, Moscow, Russia}
\affiliation{Moscow State University, Moscow, Russia}
\affiliation{Institute for High Energy Physics, Protvino, Russia}
\affiliation{Petersburg Nuclear Physics Institute, St. Petersburg, Russia}
\affiliation{Instituci\'{o} Catalana de Recerca i Estudis Avan\c{c}ats (ICREA) and Institut de F\'{i}sica d'Altes Energies (IFAE), Barcelona, Spain}
\affiliation{Uppsala University, Uppsala, Sweden}
\affiliation{Lancaster University, Lancaster LA1 4YB, United Kingdom}
\affiliation{Imperial College London, London SW7 2AZ, United Kingdom}
\affiliation{The University of Manchester, Manchester M13 9PL, United Kingdom}
\affiliation{University of Arizona, Tucson, Arizona 85721, USA}
\affiliation{University of California Riverside, Riverside, California 92521, USA}
\affiliation{Florida State University, Tallahassee, Florida 32306, USA}
\affiliation{Fermi National Accelerator Laboratory, Batavia, Illinois 60510, USA}
\affiliation{University of Illinois at Chicago, Chicago, Illinois 60607, USA}
\affiliation{Northern Illinois University, DeKalb, Illinois 60115, USA}
\affiliation{Northwestern University, Evanston, Illinois 60208, USA}
\affiliation{Indiana University, Bloomington, Indiana 47405, USA}
\affiliation{Purdue University Calumet, Hammond, Indiana 46323, USA}
\affiliation{University of Notre Dame, Notre Dame, Indiana 46556, USA}
\affiliation{Iowa State University, Ames, Iowa 50011, USA}
\affiliation{University of Kansas, Lawrence, Kansas 66045, USA}
\affiliation{Kansas State University, Manhattan, Kansas 66506, USA}
\affiliation{Louisiana Tech University, Ruston, Louisiana 71272, USA}
\affiliation{Northeastern University, Boston, Massachusetts 02115, USA}
\affiliation{University of Michigan, Ann Arbor, Michigan 48109, USA}
\affiliation{Michigan State University, East Lansing, Michigan 48824, USA}
\affiliation{University of Mississippi, University, Mississippi 38677, USA}
\affiliation{University of Nebraska, Lincoln, Nebraska 68588, USA}
\affiliation{Rutgers University, Piscataway, New Jersey 08855, USA}
\affiliation{Princeton University, Princeton, New Jersey 08544, USA}
\affiliation{State University of New York, Buffalo, New York 14260, USA}
\affiliation{University of Rochester, Rochester, New York 14627, USA}
\affiliation{State University of New York, Stony Brook, New York 11794, USA}
\affiliation{Brookhaven National Laboratory, Upton, New York 11973, USA}
\affiliation{Langston University, Langston, Oklahoma 73050, USA}
\affiliation{University of Oklahoma, Norman, Oklahoma 73019, USA}
\affiliation{Oklahoma State University, Stillwater, Oklahoma 74078, USA}
\affiliation{Brown University, Providence, Rhode Island 02912, USA}
\affiliation{University of Texas, Arlington, Texas 76019, USA}
\affiliation{Southern Methodist University, Dallas, Texas 75275, USA}
\affiliation{Rice University, Houston, Texas 77005, USA}
\affiliation{University of Virginia, Charlottesville, Virginia 22904, USA}
\affiliation{University of Washington, Seattle, Washington 98195, USA}
\author{V.M.~Abazov} \affiliation{Joint Institute for Nuclear Research, Dubna, Russia}
\author{B.~Abbott} \affiliation{University of Oklahoma, Norman, Oklahoma 73019, USA}
\author{B.S.~Acharya} \affiliation{Tata Institute of Fundamental Research, Mumbai, India}
\author{M.~Adams} \affiliation{University of Illinois at Chicago, Chicago, Illinois 60607, USA}
\author{T.~Adams} \affiliation{Florida State University, Tallahassee, Florida 32306, USA}
\author{G.D.~Alexeev} \affiliation{Joint Institute for Nuclear Research, Dubna, Russia}
\author{G.~Alkhazov} \affiliation{Petersburg Nuclear Physics Institute, St. Petersburg, Russia}
\author{A.~Alton$^{a}$} \affiliation{University of Michigan, Ann Arbor, Michigan 48109, USA}
\author{A.~Askew} \affiliation{Florida State University, Tallahassee, Florida 32306, USA}
\author{S.~Atkins} \affiliation{Louisiana Tech University, Ruston, Louisiana 71272, USA}
% ...
%
%
\collaboration{The D0 Collaboration\footnote{with visitors from
$^{a}$Augustana College, Sioux Falls, SD, USA,
$^{b}$The University of Liverpool, Liverpool, UK,
$^{c}$UPIITA-IPN, Mexico City, Mexico,
$^{d}$DESY, Hamburg, Germany,
$^{e}$SLAC, Menlo Park, CA, USA,
$^{f}$University College London, London, UK,
$^{g}$Centro de Investigacion en Computacion - IPN, Mexico City, Mexico,
$^{h}$ECFM, Universidad Autonoma de Sinaloa, Culiac\'an, Mexico
and
$^{i}$Universidade Estadual Paulista, S\~ao Paulo, Brazil.
}} \noaffiliation
\vskip 0.25cm

  \date{August 27, 2012}
  
  \begin{abstract} 
    We present final searches of the anomalous ${\gamma}WW$ and $ZWW$ 
    trilinear gauge boson couplings from $WW$ and $WZ$ production 
    using lepton plus dijet final states and a combination with results 
    from $W\gamma$, $WW$, and $WZ$ production with leptonic final states.  
    The analyzed data correspond to up to $8.6$~fb$^{-1}$ of 
    integrated luminosity collected by the D0 detector in $p\bar{p}$ 
    collisions at $\sqrt{s}=1.96$~TeV.  We set the most stringent 
    limits at a hadron collider to date assuming two different relations 
    between the anomalous coupling parameters $\Delta\kappa_\gamma$, 
    $\lambda$, and $\Delta g_1^Z$ for a cutoff energy scale 
    $\Lambda=2$~TeV.  The combined 68\% C.L. limits are 
    $-0.057<\Delta\kappa_\gamma<0.154$, $-0.015<\lambda<0.028$, 
    and $-0.008<\Delta g_1^Z<0.054$ for the LEP parameterization, and 
    $-0.007<\Delta\kappa<0.081$ and $-0.017<\lambda<0.028$ for the equal 
    couplings parameterization.  We also present the most stringent 
    limits of the $W$ boson magnetic dipole and electric quadrupole moments. 
  \end{abstract}
  
  \maketitle 
  
  In the standard model (SM), the neutral vector bosons, $\gamma$ 
  and $Z$, do not interact among themselves, while the charged 
  vector bosons, $W^{\pm}$, couple with the neutral ones and among 
  themselves through trilinear and quartic gauge interactions.  The 
  most general $\gamma{WW}$ and $ZWW$ interactions can be described 
  using a Lorentz invariant effective Lagrangian that contains fourteen 
  dimensionless couplings, seven each for $\gamma WW$ and 
  $ZWW$~\cite{bib:lagrangian,bib:HWZ}.  Assuming $C$ (charge) and 
  $P$ (parity) conservation and electromagnetic gauge invariance, i.e. 
  $g^\gamma_1=1$ where $g^\gamma_1$ is the $C$ and $P$ conserving trilinear 
  gauge boson coupling, reduces the number of independent couplings to 
  five, and the Lagrangian terms take the form:
  \begin{equation} \begin{array}{ccl} \frac{\mathcal{L}_{VWW}}{g_{VWW}} & = & i g_{1}^{V}
  (W_{\mu\nu}^{\dag}W^{\mu}V^{\nu} - W_{\mu}^{\dag}V_{\nu}W^{\mu\nu}) \\ & +
  & i{\kappa}_{V}W_{\mu}^{\dag}W_{\nu}V^{\mu\nu} +
  i\frac{\lambda_{V}}{M_{W}^{2}}
  W_{\lambda\mu}^{\dag}W_{\nu}^{\mu}V^{\nu\lambda},
  \label{eq:eq-lag} \end{array}{} \end{equation}
  \noindent where $W^{\mu}$ denotes the $W$ boson field, $V^\mu$ is 
  either the photon ($V=\gamma$) or the $Z$ boson ($V=Z$) field, 
  $W_{\mu\nu}=\partial_{\mu}W_{\nu}-\partial_{\nu}W_{\mu}$, $V_{\mu\nu} = 
  \partial_{\mu}V_{\nu}-\partial_{\nu}V_{\mu}$, and $M_W$ is the mass 
  of the $W$ boson.  The global coupling parameters $g_{VWW}$ are 
  $g_{\gamma WW}=-e$ and $g_{ZWW}=-e\,{\rm cot}\theta _W$, as in the SM, where 
  $e$ and $\theta_W$ are the magnitude of the electron 
  charge and the weak mixing angle, respectively.  In the SM, the five 
  remaining couplings are $\lambda_\gamma=\lambda_Z=0$ and 
  $g_1^Z=\kappa_{\gamma}=\kappa_Z=1$.  Any deviation of these couplings 
  from their predicted SM values would be an indication for new 
  physics~\cite{bib:strong} and could provide information on a mechanism 
  for electroweak symmetry breaking.  These deviations are denoted as the 
  anomalous trilinear gauge couplings (ATGCs) $\Delta\kappa_V$ and 
  $\Delta g_1^Z$ defined as  $\kappa_V-1$ and $g_1^Z-1$, respectively.  The 
  $W$ boson magnetic dipole moment, $\mu_W$, and electric quadrupole moment, 
  $q_W$, are related to the coupling parameters by:
  \begin{equation}
    \mu_W = \frac{e}{2M_W} (1 + \kappa_\gamma + \lambda_\gamma),~ 
    q_W = - \frac{e}{M^2_W} (\kappa_\gamma - \lambda_\gamma). 
    \label{eq:moments}
  \end{equation}
  
  If the coupling parameters have non-SM values, the amplitudes for gauge 
  boson pair production may grow with energy, eventually violating tree-level 
  unitarity.  Unitarity violation can be controlled by parameterizing the 
  ATGCs as dipole form factors with a cutoff energy scale, $\Lambda$.  The ATGCs 
  then take the form $a(\hat{s}) = a_0 / (1 + \hat{s}/\Lambda^{2})^{2}$ 
  in which $\sqrt{\hat{s}}$ is the center-of-mass energy of the colliding partons 
  and $a_0$ is the coupling value in the limit $\hat{s} \rightarrow 0$~\cite{bib:newref}.  
  The quantity $\Lambda$ is interpreted as the energy scale where the 
  new phenomenon responsible for the ATGCs is directly observable.  At the 
  Tevatron the cutoff scale $\Lambda=2$~TeV is chosen such that the unitarity 
  limits are close to, but not tighter than, the coupling limits set by data. 
  
  We assume two scenarios for studying the ATGCs.  The parameterization 
  used by the LEP experiments~\cite{bib:leppar} (we refer to this as the LEP 
  parameterization) assumes the following relation between the ATGCs:
  \begin{equation}
    \Delta\kappa_Z = \Delta g_1^Z - \Delta\kappa_\gamma \cdot \tan^2\theta_W, \text{ and } 
    \lambda_Z = \lambda_\gamma = \lambda.
    \label{eq:lepparam}
  \end{equation}
  In the equal couplings scenario~\cite{bib:HWZ}, the $\gamma WW$ and the 
  $ZWW$ couplings are set equal to each other and are sensitive to interference 
  effects between the photon and $Z$-exchange diagrams in $WW$ production.  
  Electromagnetic gauge invariance requires that $\Delta g_1^Z=\Delta g_1^\gamma=0$ 
  and 
  \begin{equation}
    \Delta\kappa_Z = \Delta\kappa_\gamma = \Delta\kappa \text{ and } 
    \lambda_Z = \lambda_\gamma = \lambda.
    \label{eq:eqparam}
  \end{equation}
  \noindent In the following analyses, we consider these two scenarios and set 
  limits on $\Delta\kappa_{\gamma}$, $\lambda$, and $\Delta g_1^Z$ 
  assuming the relations above with $\Lambda=2$~TeV.
  
  Previously published combined limits on ATGCs at the Tevatron come 
  from the D0 Collaboration from a combination of $W\gamma\rightarrow\ell\nu\gamma$, 
  $WW\rightarrow\ell\nu\ell\nu$, $WW+WZ\rightarrow\ell\nu{jj}$ and 
  $WZ\rightarrow\ell\nu{ee}$ channels ($j$ is a jet, $\ell$ is an electron, 
  $e$, or muon, $\mu$, and $\nu$ is a neutrino) with integrated luminosity, 
  $\cal{L}$, up to 100~pb$^{-1}$~\cite{bib:run1prd}, and from the CDF 
  Collaboration from a combination of $WW+WZ\rightarrow\ell\nu{jj}$ and 
  $W\gamma\rightarrow\ell\nu\gamma$ channels with 
  $\cal{L}\approx$~350~pb$^{-1}$~\cite{bib:cdfresult}.  The LEP experiments 
  published ATGC limits analyzing primarily $WW$ 
  production~\cite{bib:aleph,bib:opal,bib:l3,bib:delphi} while the CMS and 
  ATLAS experiments at the LHC $pp$ collider have published limits on 
  $\gamma{WW}/ZWW$ couplings from individual $W\gamma$, $WW$ and $WZ$ 
  final states~\cite{bib:ATLAS,bib:CMS}.  
  
  In this Letter, we measure the coupling parameters at the $\gamma WW$ 
  and $ZWW$ trilinear vertices through the study of gauge boson pair 
  production.  While the $WZ$ ($W\gamma$) final states are exclusively 
  produced via the $ZWW$ ($\gamma{WW}$) couplings, the $WW$ final state 
  can be produced through both $\gamma{WW}$ and $ZWW$ couplings.  First, 
  we present new 4.3~fb$^{-1}$ ATGC results from $WW+WZ\rightarrow\ell\nu{jj}$ 
  production and new 8.6~fb$^{-1}$ ATGC results from $WZ\rightarrow\ell\nu{\ell\ell}$ 
  production where a $W$ boson decays leptonically and the other boson decays 
  into a dijet or dilepton pair.  These results are then combined with 
  previously published ATGC measurements from 
  $W\gamma\to\ell\nu\gamma$~\cite{bib:wgamma1,bib:wgamma2}, 
  $WW\rightarrow\ell\nu\ell\nu$~\cite{bib:xsec1} and 
  $WW+WZ\rightarrow\ell\nu{jj}$~\cite{bib:lvjj1} production which analyzed 
  4.9~fb$^{-1}$, 1.0~fb$^{-1}$ and 1.1~fb$^{-1}$ of integrated luminosity, 
  respectively.  The 1.1~fb$^{-1}$ of integrated luminosity used in the 
  previous analysis of $\ell\nu{jj}$ final states is independent from the 
  data analyzed in this Letter.  Each measurement used data collected by 
  the D0 detector~\cite{bib:dzeronim} from $p\bar{p}$ collisions at 
  $\sqrt{s}=1.96$~TeV delivered by the Fermilab Tevatron Collider.  
  
  The D0 detector is a general purpose collider detector consisting of a 
  central tracking system located within a 2~T superconducting solenoidal 
  magnet, a hermetic liquid-argon and uranium calorimeter~\cite{bib:run1det}, 
  and an outer muon system~\cite{bib:run2muon} surrounding 1.8~T iron toroids.  
  Details on the reconstruction and identification criteria for electrons, 
  muons, jets, and missing transverse energy, \met, and for selection of 
  $W\gamma\to\ell\nu\gamma$, $WW\rightarrow\ell\nu\ell\nu$, 
  $WW+WZ\rightarrow\ell\nu{jj}$, and $WZ\rightarrow\ell\nu\ell\ell$ final 
  states can be found 
  elsewhere~\cite{bib:wgamma1,bib:wgamma2,bib:xsec1,bib:lvjj1,bib:ourWjets,bib:mika}. 
  
  The analysis of $WW+WZ\rightarrow\ell\nu{jj}$ final states extends a 
  previous D0 analysis of 4.3~fb$^{-1}$ of integrated luminosity which 
  measured the $WW$ and $WZ$ cross sections~\cite{bib:ourWjets}.  To 
  select $WW+WZ\rightarrow\ell\nu{jj}$ candidates, we require a single 
  reconstructed electron (muon) with transverse momentum $p_T>15~(20)$~GeV 
  and pseudorapidity $|\eta|<1.1\ (2.0)$~\cite{bib:def}, $\met>20$~GeV, 
  two or three reconstructed jets with $p_T>20$~GeV and $|\eta|<2.5$, and 
  the $W$ transverse mass~\cite{bib:smithUA1}, $M_T^{\ell\nu}~(\rm{GeV})>40-0.5$\met.  
  The reconstructed transverse momentum of the two most energetic jets 
  ($p_T^{jj}$) of selected $\ell\nu{jj}$ candidates is used to search 
  for ATGCs.  In order to maximize the sensitivity to ATGCs, only candidate 
  events within a dijet invariant mass in the range of $55<M_{jj}<110$~GeV 
  are studied.
  
  The ATGC analysis of $WZ\rightarrow\ell\nu{\ell\ell}$ final states builds 
  upon a previous D0 measurement of the $WZ$ cross section~\cite{bib:mika} 
  with 8.6~fb$^{-1}$ of integrated luminosity and uses the reconstructed 
  transverse momentum of the two leptons ($p_T^{\ell\ell}$) originating 
  from the $Z$ boson.  To select $WZ\rightarrow\ell\nu\ell\ell$ candidates, 
  we require $\met>20$~GeV, at least two oppositely charged electrons (muons) 
  with $|\eta|<3.0\ (2.0)$, $p_T^{1}>20~(15)$~GeV and $p_T^{2}>15~(10)$~GeV, 
  and with an invariant mass $60<M_{\ell\ell}<120$~GeV.  An additional electron 
  or muon is required to have $p_T>15$~GeV.  In the case of three like-flavor 
  leptons, the oppositely charged lepton pair with $M_{\ell\ell}$ more consistent 
  with the $Z$ boson mass is assigned to the $Z$ decay provided that at least 
  one of the two leptons has $p_T>25$~GeV.  Otherwise the event is rejected.
  
  The SM $WW+WZ\rightarrow\ell\nu{jj}$ and $WZ\rightarrow\ell\nu{\ell\ell}$ 
  production and most of the other background processes are modeled using Monte 
  Carlo (MC) simulation.  In $\ell\nu{jj}$ production the dominant background 
  is due to the production of a vector boson ($V=W,Z$) in association with 
  jets from light or heavy flavor parton production followed by the production 
  of singe top quarks or top quark pairs.  These backgrounds are modeled by 
  MC simulation, while the multijet background is determined from data.  In 
  $\ell\nu\ell\ell$ production the dominant 
  $Z/\gamma^{*}\rightarrow\ell\ell$, $ZZ$ and $Z\gamma$ backgrounds are 
  modeled with MC.  Detailed information about the background modeling can 
  be found elsewhere~\cite{bib:ourWjets,bib:mika}.  The SM 
  $WW+WZ\rightarrow\ell\nu{jj}$ and $WZ\rightarrow\ell\nu{\ell\ell}$ events 
  are generated with {\sc pythia}~\cite{bib:PYTHIA} using 
  \textsc{CTEQ6L1} parton distribution functions (PDFs)~\cite{bib:CTEQ6}.  
  {\sc pythia} is a leading order (LO) generator, therefore we correct the 
  event kinematics and the acceptance of $\ell\nu{jj}$ events for 
  next-to-LO (NLO) and resummation effects.  To derive this correction we 
  use \textsc{mc@nlo}~\cite{bib:mc@nlo} with \textsc{CTEQ6.1M} PDFs 
  interfaced to {\sc herwig}~\cite{bib:herwig} for parton showering and 
  hadronization.  Comparing {\sc pythia} to \textsc{mc@nlo} kinematics 
  at the generator level after final state radiation, we parameterize
  a two-dimensional correction matrix in the $p_T$ of the diboson system 
  and that of the highest-$p_{T}$ boson, and use it to reweight the {\sc pythia} 
  $\ell\nu{jj}$ events.  The event yields for the $WW\rightarrow\ell\nu{jj}$ 
  and $WZ\rightarrow\ell\nu{jj}$ production are normalized to the SM NLO 
  cross sections of $\sigma(WW)=11.7\pm 0.8$~pb and $\sigma(WZ)=3.5\pm 0.3$~pb 
  calculated with {\sc mcfm}~\cite{bib:MCFM} using \textsc{MSTW2008} PDFs.  
  The above procedure is designed to give NLO predictions at the detector level 
  for the SM contributions to the diboson processes.  The $WZ\rightarrow\ell\nu\ell\ell$ 
  events are also generated using {\sc pythia} with \textsc{CTEQ6L1} PDFs 
  and thus also need to be corrected as a function of diboson $p_{T}$ to 
  match predictions from the NLO event generator {\sc powheg}~\cite{bib:POWHEG}.  
  The event yields for $WZ\rightarrow\ell\nu{\ell\ell}$ production are 
  normalized to the SM NLO cross section of $\sigma(WZ)=3.2\pm 0.1$~pb 
  calculated for the $Z$ boson invariant mass range of $60<M_{Z}<120$~GeV 
  with {\sc mcfm} and \textsc{MSTW2008} PDFs.  

  All MC events undergo a {\sc geant}-based~\cite{bib:GEANT} detector 
  simulation and are reconstructed using the same algorithms as used for 
  D0 data.  The effect of multiple $p\bar{p}$ interactions is included by 
  overlaying data events from random beam crossings on simulated events.  
  We apply corrections to the MC to account for differences with data 
  in reconstruction and identification efficiencies of leptons and jets.  
  Trigger efficiencies measured in data are applied to MC.  The 
  instantaneous luminosity profile and $z$ distribution of the $p\bar{p}$ 
  interaction vertex of each MC sample are adjusted to match those in data.  
  
  In order to extract the ATGCs, we follow a two-step procedure which 
  allows to save computing time.  We first use the {\sc geant}-based D0 event 
  simulation of diboson processes, reweighted with a SM NLO model of 
  diboson production to produce a baseline sample of simulated events 
  for comparison with data.  We then use a simulation with ATGCs to 
  create a set of predictions relative to the SM, $R\propto\sigma/\sigma_{SM}$.  
  The ratio $R$ is used to reweight the SM GEANT-based simulation to 
  reflect ATGCs.  This reweighted simulation is then compared to data 
  and used to extract possible values of the ATGCs.
  
  The effect of ATGCs is to increase the production cross section, 
  especially at high boson transverse momentum, relative to its SM prediction.  
  We therefore use the corresponding $p_T^{jj}$ and $p_T^{\ell\ell}$ 
  distributions to set the limits on ATGCs.  The SM $p_T^{jj}$ and 
  $p_T^{\ell\ell}$ distributions are reweighted with $R$ at the 
  parton level.  The reweighting method uses the ratio of matrix element 
  squared values with and without the ATGC component to predict a change 
  of the  SM event rate in the presence of ATGCs.  The basis of the 
  reweighting method is that the equation of the differential cross section, 
  which has a quadratic dependence on the anomalous couplings, can be written 
  as:
  \begin{equation}
    \begin{array}{ccl} d\sigma & \propto & |\mathcal M|^{2}dx \\ & \propto & |\mathcal
      M|_{SM}^{2}\frac{|{\mathcal M}|^{2}}{|{\mathcal M}|_{SM}^{2}}dx \\ &
      \propto & |\mathcal M|^{2}_{SM}[1+A\Delta\kappa+B(\Delta\kappa)^{2}\\ & + &
	C\lambda + D\lambda^{2}+E\Delta\kappa\lambda
	+ etc ... ]dx \\ & \propto & d\sigma_{SM}\cdot R(\Delta\kappa,\lambda,...),  
	\label{eq:expan} 
    \end{array}{} 
  \end{equation}
  \noindent where $d\sigma$ is the differential cross section that includes the 
  contribution from the ATGCs; $d\sigma_{SM}$ is the SM differential 
  cross section; $|{\mathcal M}|^{2}$ is the matrix element squared 
  in the presence of ATGCs; $|{\mathcal M}|_{SM}^{2}$ is the matrix 
  element squared in the SM; $A,~B,~C,~D$,~$E$, etc.~are reweighting 
  coefficients; and $x$ is a kinematic variable sensitive to ATGCs.  
  
  In the LEP parametrization, Eq.~(\ref{eq:expan}) is parametrized with 
  the three couplings $\Delta\kappa_{\gamma},~\lambda$, and 
  $\Delta{g_{1}^{Z}}$ and nine reweighting coefficients, $A$-$I$.  
  Thus, the weight $R$ in the LEP parametrization scenario is defined 
  as:
  \begin{equation}
    \begin{array}{ccl}
      &R&(\Delta\kappa_\gamma,\lambda,\Delta g_{1}) = 1+A\Delta\kappa_\gamma \\
      &+& B(\Delta\kappa_\gamma)^{2}+C\lambda+D\lambda^{2} \\
      &+& E\Delta g_{1}+F(\Delta g_{1})^{2}+G\Delta\kappa_\gamma\lambda \\
      &+& H\Delta\kappa_\gamma \Delta g_{1}+I\lambda\Delta g_{1},
      \label{eq:expan1} 
    \end{array}{} 
  \end{equation}
  \noindent with $\lambda=\lambda_{\gamma}=\lambda_{Z}$ and 
  $\Delta{g_{1}}=\Delta{g_{1}^{Z}}$.  
  
  In the equal couplings scenario, Eq.~(\ref{eq:expan}) is parametrized with the 
  two couplings $\Delta\kappa$ and $\lambda$ and five reweighting coefficients, 
  $A$-$E$.  In this case the weight is defined as:
  \begin{equation} \begin{array}{ccl} &R&(\Delta\kappa,\lambda) =
      1+A\Delta\kappa+B(\Delta\kappa)^{2} \\
      &+&C\lambda+D\lambda^{2}+E\Delta\kappa\lambda,
      \label{eq:expan2} 
    \end{array}{} 
  \end{equation}
  \noindent with $\Delta\kappa=\Delta\kappa_{\gamma}=\Delta\kappa_{Z}$ and 
  $\lambda=\lambda_{\gamma}=\lambda_{Z}$.  Depending on the number 
  of reweighting coefficients, a system of the same number of equations 
  allows us to calculate their values for each event.  Then for any 
  ATGC combination we can calculate $R$ and apply it to the SM 
  distribution to describe that kinematic distribution in the 
  presence of the chosen non-SM TGC.  We first calculate $R_{i}$ 
  ($i=1-5$ for the equal couplings scenario and $i=1-9$ for the	
  LEP parameterization) with a fixed set of ATGCs using a LO 
  prediction from the {\sc mcfm} generator (with \textsc{CTEQ6L1} 
  PDFs).  Therefore each {\sc mcfm} event is assigned a value 
  of $|\mathcal M|^{2}_{SM}$ and a set of $|\mathcal M|^{2}$ 
  values for $\Delta\kappa_{\gamma}=\pm1$, $\lambda=\pm1$, 
  $\Delta{g_{1}^{Z}}=\pm1$, $\Delta\kappa_{\gamma}\lambda=+1$,
  $\Delta\kappa_{\gamma}\Delta{g_{1}^{Z}}=+1$, and 
  $\lambda\Delta{g_{1}^{Z}}=+1$.  For every bin $X$ in the 
  multidimensional phase space defined by different kinematic 
  distributions, the ratio R is calculated as:
  \begin{equation}
    R_{i;X} = \frac{\sum_{j}{|\mathcal M_{i,j}|^{2}_{X}}}
    {\sum_{j}{|\mathcal M^{SM}_{i,j}|^{2}_{X}}},
    \label{eq:norm}
  \end{equation}
  \noindent where $j$ indicates the event number in bin $X$, and $i$ 
  is any of nine (five) ATGC combinations in the LEP parameterization 
  (equal couplings scenario).  The multidimensional phase space for the 
  $WW\rightarrow\ell\nu{jj}$ events is defined by a set of kinematic 
  variables at generator level, namely the transverse momentum ($p_{T}$) 
  of the $q\bar{q}$ system, $p_{T}^{q\bar{q}}$, $p_{T}$ of the leading 
  parton, $p_{T}$ of the trailing parton, $p_{T}$ of the neutrino, $p_{T}$ 
  of the charged lepton, and the invariant mass of the $q\bar{q}$ system.  
  For $WZ\rightarrow\ell\nu{jj}$ events, we use the $p_{T}$ distribution 
  of the quark, the $p_{T}$ distribution of the anti-quark, $p_{T}^{q\bar{q}}$, 
  $p_{T}$ of the neutrino, $p_{T}$ of the charged lepton, and the invariant 
  mass of the $q\bar{q}$ system.  For $WZ\rightarrow\ell\nu{\ell\ell}$ 
  events, $X$ is defined by the transverse momentum of the dilepton system, 
  $p_{T}^{\ell\ell}$, where both leptons originate from the $Z$ boson,
  $p_{T}$ of the leading and the trailing leptons originating from the 
  $Z$ boson, $p_{T}$ of the lepton originating from the $W$ boson, and 
  $p_{T}$ of the neutrino.  
  
  When searching for ATGCs in the LEP parametrization, we vary two of the 
  three couplings at a time, leaving the third coupling fixed to its 
  SM value.  This gives the three two-parameter combinations 
  ($\Delta\kappa_{\gamma},\lambda$), ($\Delta\kappa_{\gamma},
  \Delta g_{1}^{Z}$), and ($\lambda,\Delta g_{1}^{Z}$).  For the equal 
  couplings scenario there is only the ($\Delta\kappa,\lambda$) combination.  
  For a given pair of ATGC values, each SM event is weighted at the 
  generator level by the appropriate weight $R_{i;X}$ and all the weights 
  in a reconstructed $p_{T}^{jj}$ (or $p_{T}^{\ell\ell}$) bin are summed.  
  Such reweighted SM distributions are compared to data to determine which 
  ATGCs are most consistent with observation.  Kinematic distributions 
  in $W\gamma\rightarrow\ell\nu\gamma$ and $WW\rightarrow\ell\nu\ell\nu$ 
  production sensitive to ATGCs are the $E_{T}$ of the photon, $E_{T}^{\gamma}$, 
  and $p_{T}$ distributions of the two leptons, respectively.  The 
  effects of ATGCs on the $E_{T}^{\gamma}$ distribution are modeled 
  using simulated events from the BHO generator~\cite{bib:baur} which 
  undergo {\sc geant}-based D0 detector simulation.  In case of 
  $\ell\nu\ell\nu$ final states, the ATGCs effects on $p_{T}$ distributions 
  of the two leptons are simulated using the HWZ generator~\cite{bib:HWZ} 
  and passed through a parameterized simulation of the D0 detector that 
  is tuned to data. 
  
  In order to verify the derived reweighting parameters, we calculate the weights 
  $R_{i;X}$ for different $\Delta\kappa,~\lambda$, and/or $\Delta{g_{1}^{Z}}$ 
  values, apply the reweighting coefficients and compare reweighted 
  $p_{T}$ shapes at the generator level to those predicted by {\sc mcfm}.  
  This procedure is also repeated after applying generator level 
  selection cuts similar to those at the reconstructed level to check 
  that the (acceptance~$\times$~efficiency) for reconstructed events is 
  reasonably modeled by this reweighting method.  The agreement in the 
  shape and normalization of the $p_{T}^{q\bar{q}}$ and $p_{T}^{\ell\ell}$ 
  distributions used for the ATGC measurement is within 5\% of the 
  {\sc mcfm} predictions and thus a conservative systematic 
  uncertainty of 5\% has been assigned to the reweighting method.  
  
  In the ATGC analysis of $\ell\nu{jj}$ final states, we consider 
  two classes of systematic uncertainties:  those related 
  to the overall normalization and efficiencies of the various 
  contributing physical processes, and uncertainties that, when 
  propagated through the analysis, impact the shape of the dijet 
  $p_{T}$ distribution.  We determine the dependence of the dijet 
  $p_{T}$ distribution on these uncertainties by varying each 
  parameter by its uncertainty ($\pm1$~standard deviation) and 
  re-evaluating the shape of the dijet $p_{T}$ distribution.  The 
  uncertainties with the largest impact are those related to 
  background cross sections ($6.3-20\%$), integrated luminosity 
  ($6.1\%$), the jet energy scale ($3-9\%$) and the jet energy 
  resolution ($1-10\%$) although the analysis of the $\ell\nu{jj}$ 
  final states is fully dominated by statistical uncertainty.  In 
  the analysis of $\ell\nu\ell\ell$ final states the most important 
  systematic uncertainties arise from the diboson $p_{T}$ modeling 
  ($0.1-0.4\%$), the lepton/jet energy scale ($0.2-6.0\%$), and the 
  mis-modeling of lepton/jet resolution ($1\%$).  However, the systematic uncertainties 
  are negligible compared to statistical uncertainties.  Similarly, 
  the $\ell\nu\ell\nu$ final states are mainly affected by statistical 
  uncertainty while the systematic uncertainties arise from the 
  background modeling ($<7\%$), integrated luminosity ($6.1\%$), lepton 
  identification and trigger efficiencies ($<3\%$).  In the analysis 
  of $\ell\nu\gamma$ final states systematic uncertainties due to 
  integrated luminosity ($6.1\%$), lepton and photon identification 
  ($1-5\%$), background modeling ($1-10\%$) and theoretical predictions 
  on the production cross sections ($3-6\%$) dominate the total uncertainty.

  \begin{figure*}[tbp] 
    \begin{centering}
      \includegraphics[width=8.0cm]{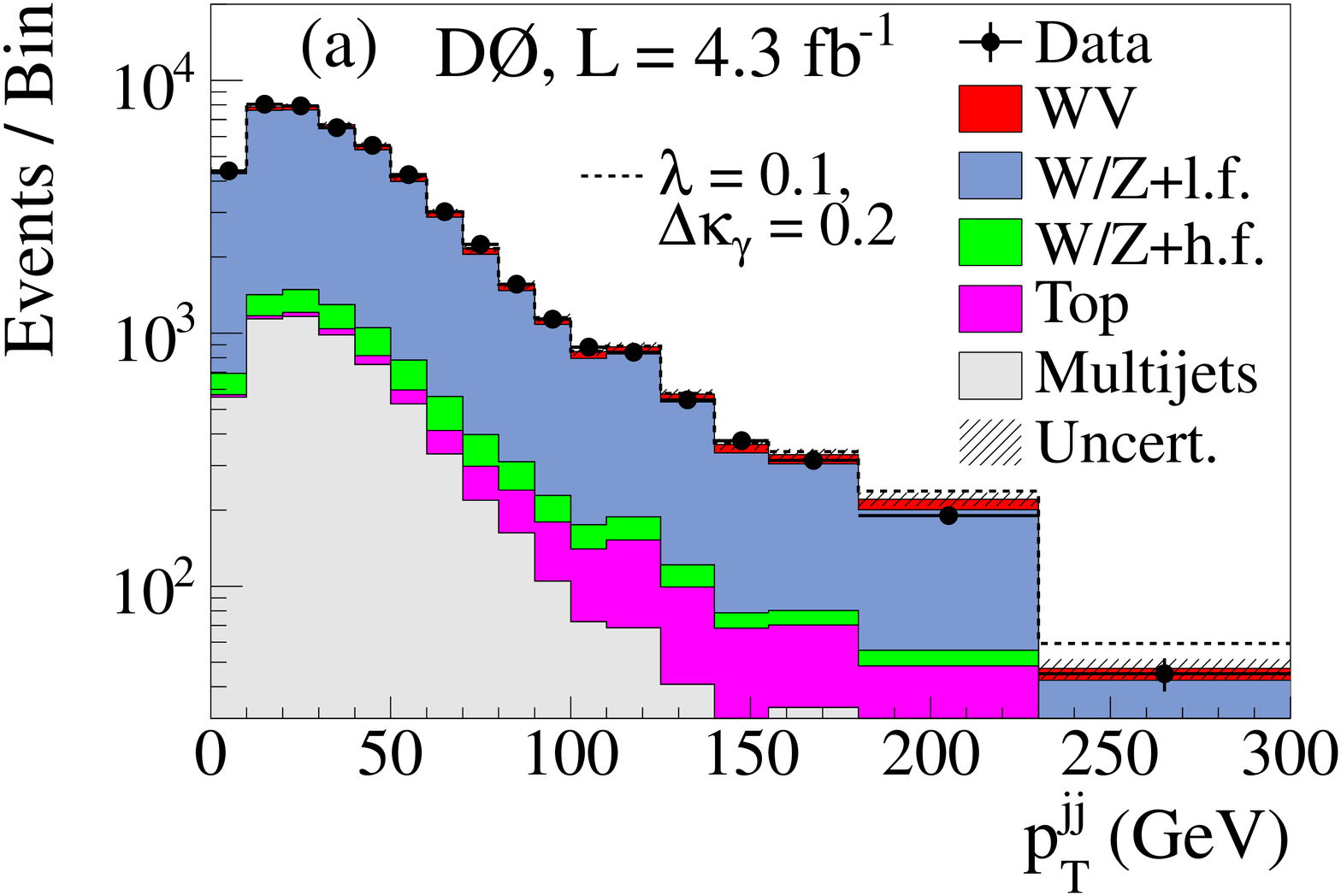}
      \includegraphics[width=8.0cm]{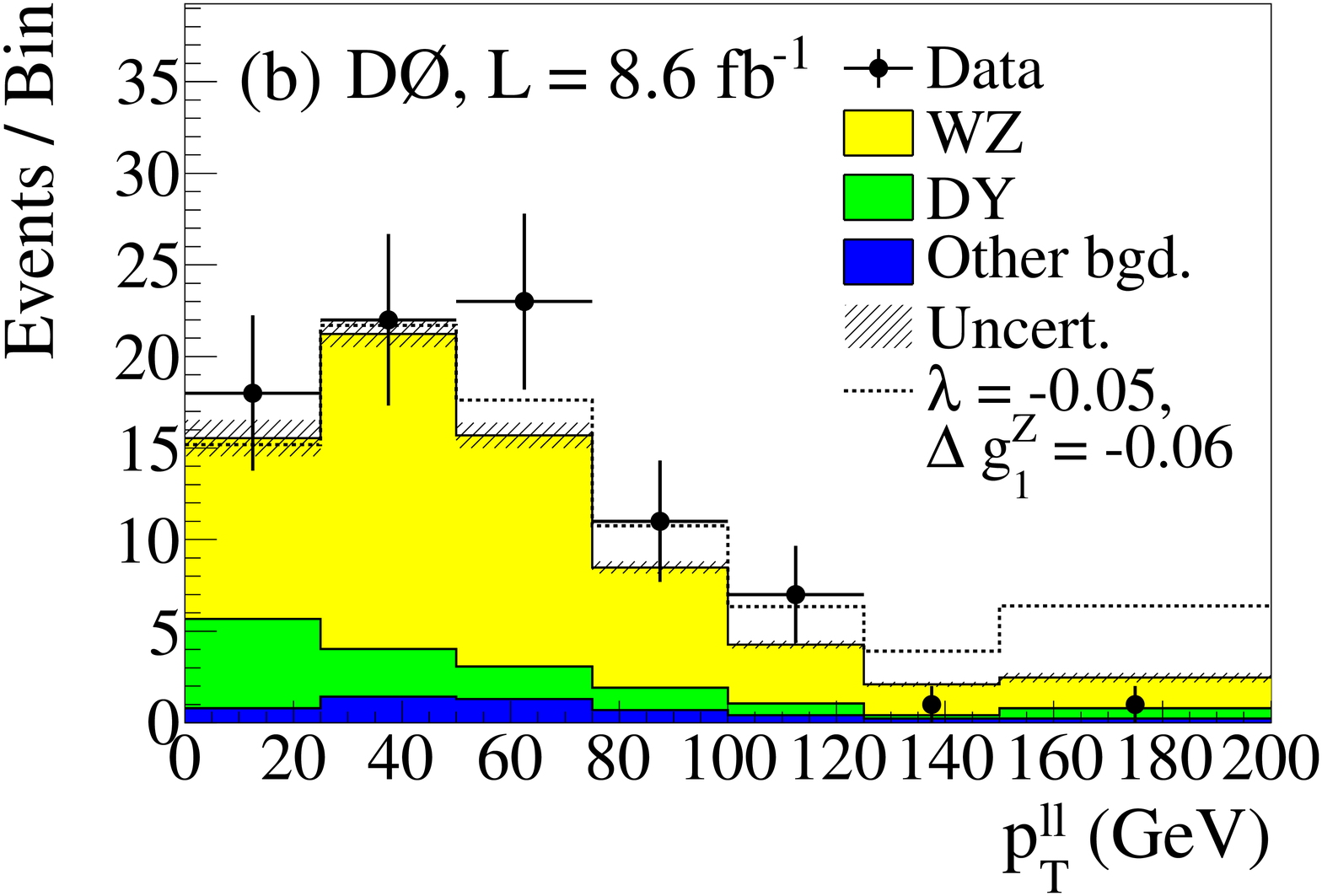}
      \caption{(color online) (a) The $p_{T}^{jj}$ distribution summed over electron 
      and muon channels from $WW+WZ\rightarrow\ell\nu{jj}$ ($l=\mu,e$) production for 
      data and SM MC predictions (``l.f.'' denotes light partons such as $u$, $d$, $s$ 
      or gluon, and ``h.f.'' denotes heavy-flavor such as $c$ or $b$).  
      Also shown are expected distributions for an ATGC model with $\Delta\kappa_{\gamma}=0.2$, 
      and $\lambda=0.1$. (b) The $p_{T}^{ll}$ distribution summed 
      over $eee$, $e\mu\mu$, $\mu{ee}$ and $\mu\mu\mu$ channels from 
      $WZ\rightarrow\ell\nu\ell\ell$ production for data, SM MC predictions and for ATGC 
      model with $\lambda=-0.05$ and $\Delta g_{1}^{Z}=-0.06$.} 
      \label{fig:kinemWWWZ} 
    \end{centering} 
  \end{figure*}

  \begin{figure*}[tbp] 
    \begin{centering}
      \includegraphics[width=8.0cm]{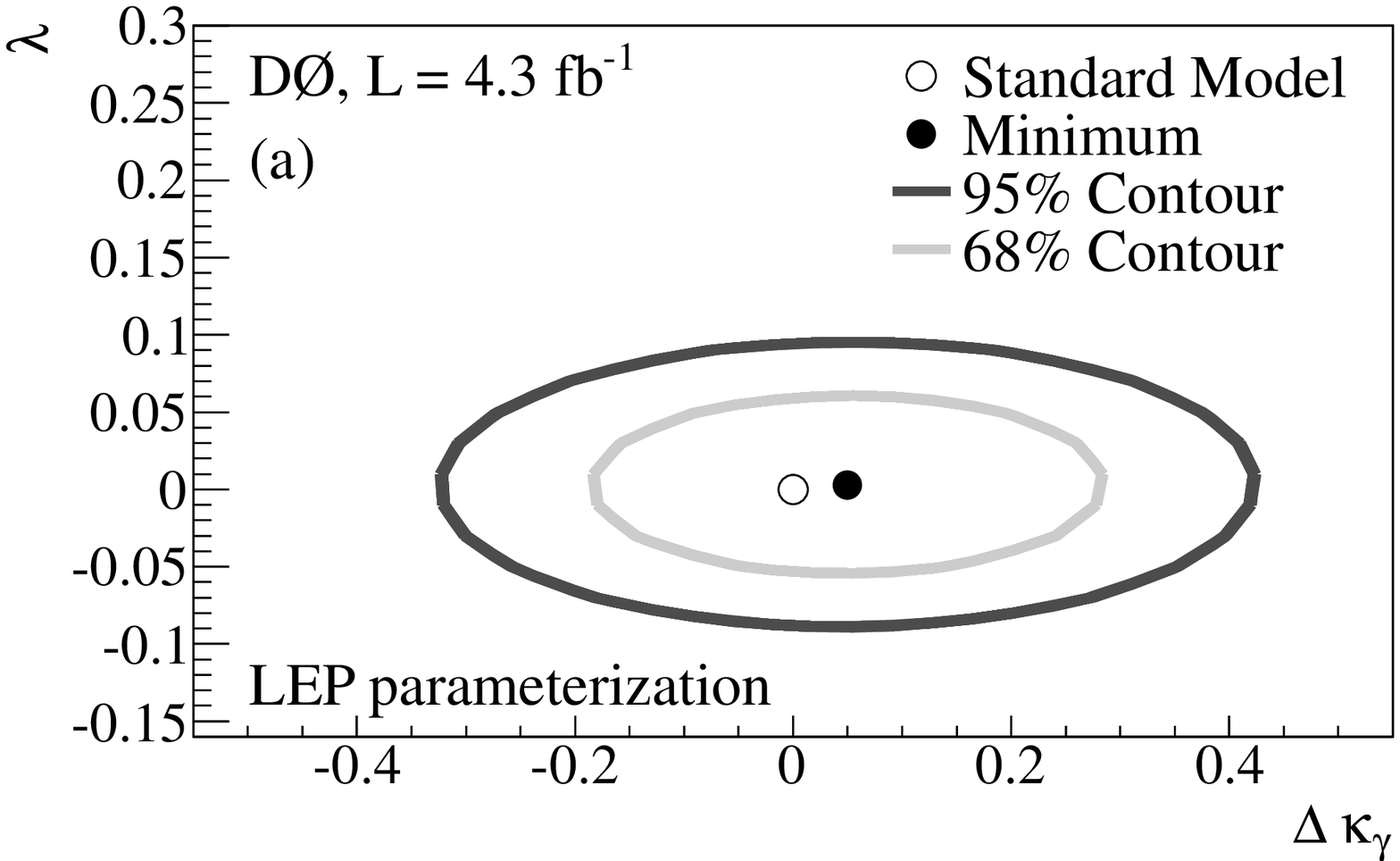}
      \includegraphics[width=8.0cm]{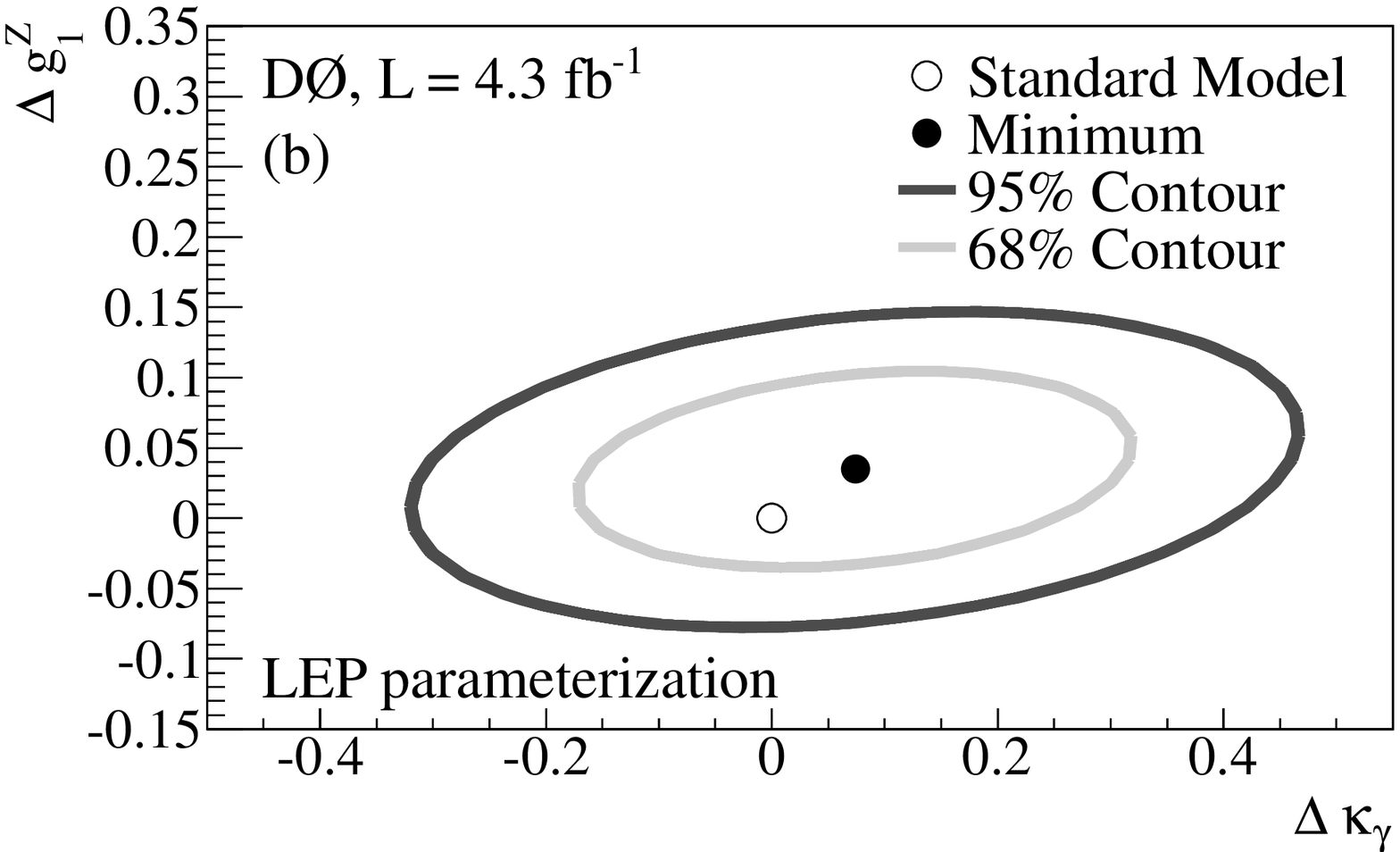} \\
      \includegraphics[width=8.0cm]{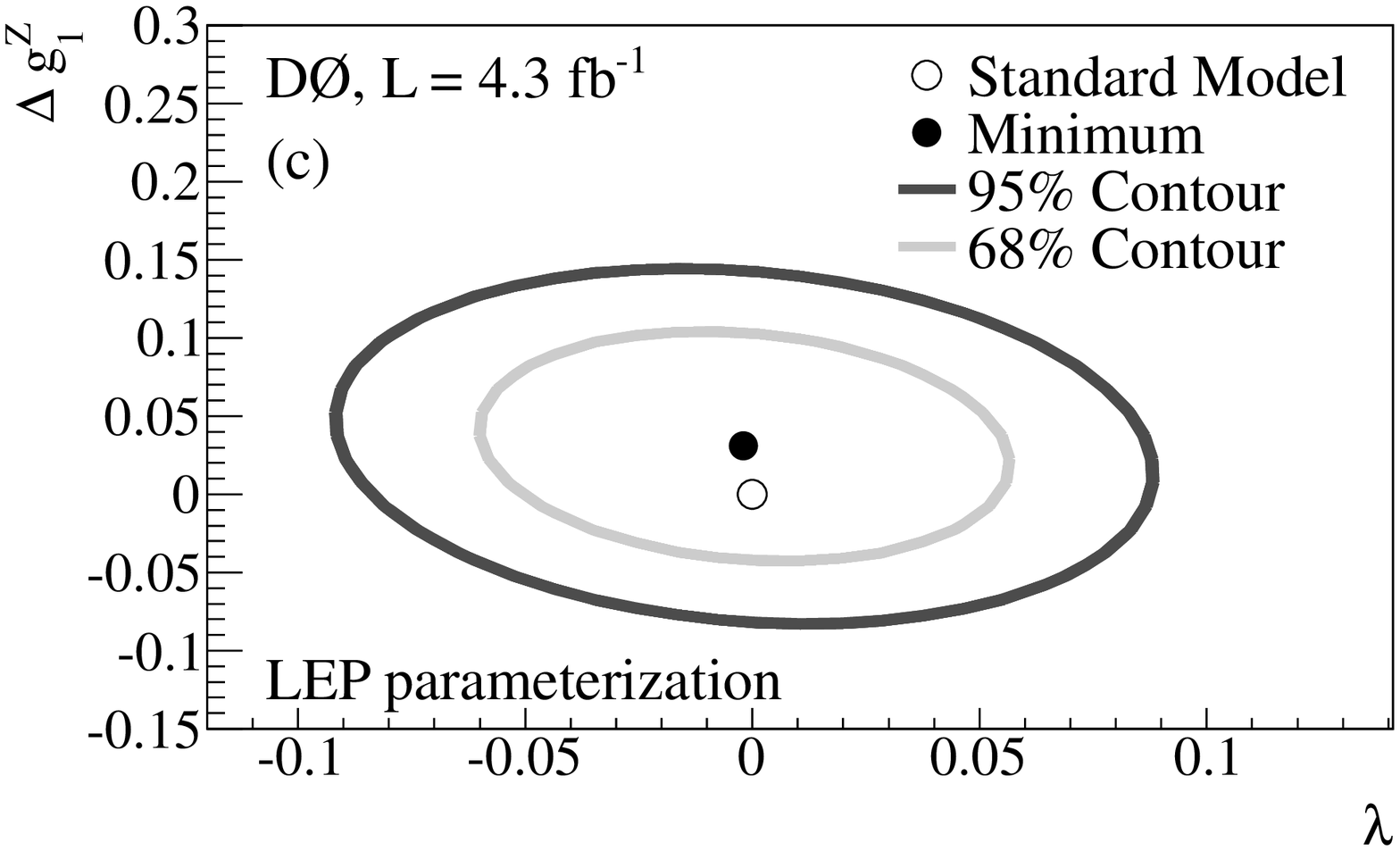}
      \includegraphics[width=8.0cm]{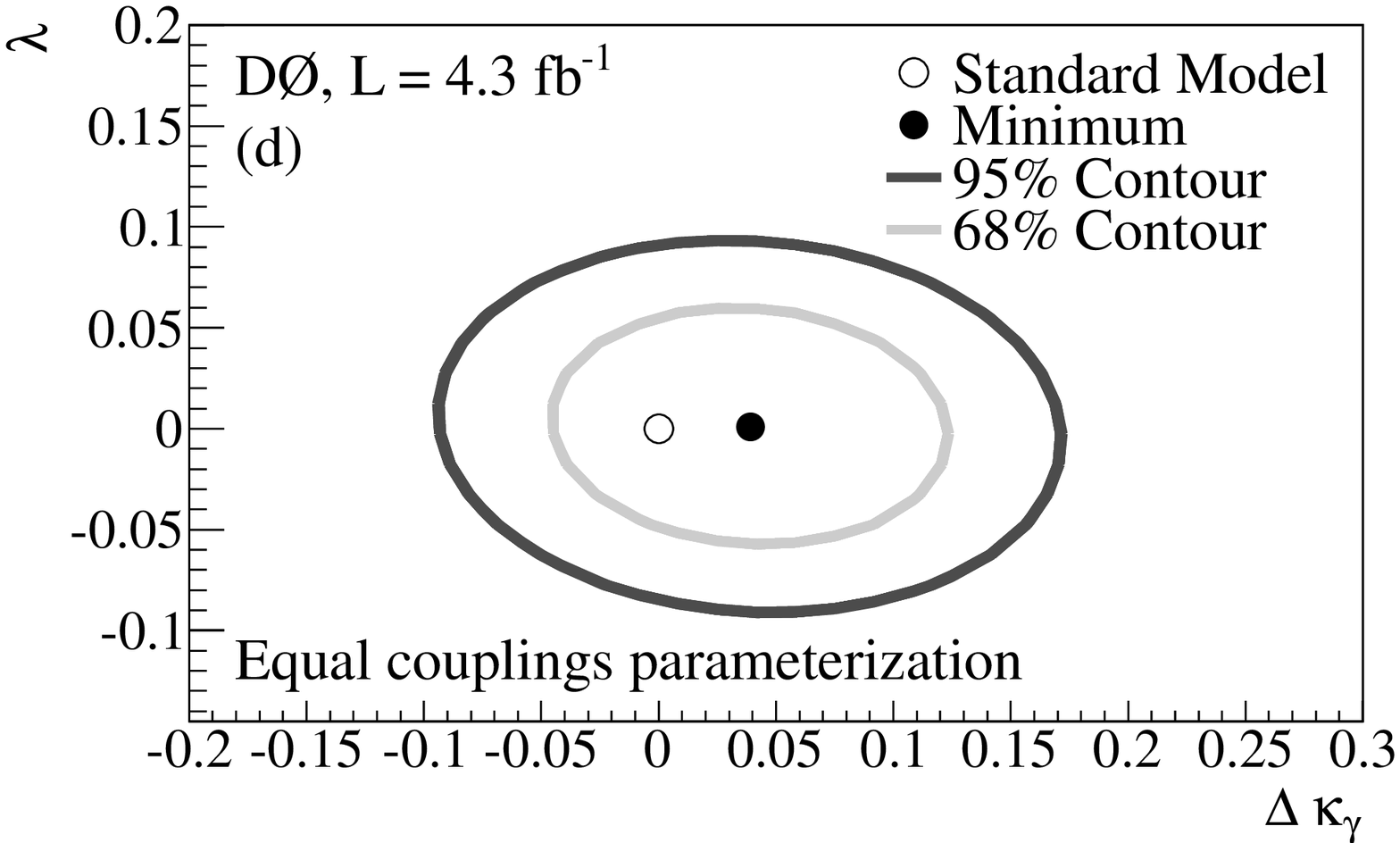} \\
      \caption{$WW+WZ\rightarrow{\ell\nu}jj$~($l=\mu,e$). 
	The 68\% and 95\% C.L. two-parameter limits on the $\gamma WW/ZWW$ 
	coupling parameters assuming the LEP (a,~b,~c) and equal 
	couplings parameterization (d) with $\Lambda=2$~TeV.  Black circles 
	indicate the most probable values of an ATGCs from the two-parameter fit.} 
      \label{fig:limit1} 
    \end{centering} 
  \end{figure*}

  \begin{figure}[tbp] 
    \begin{centering}
      \includegraphics[width=8.0cm]{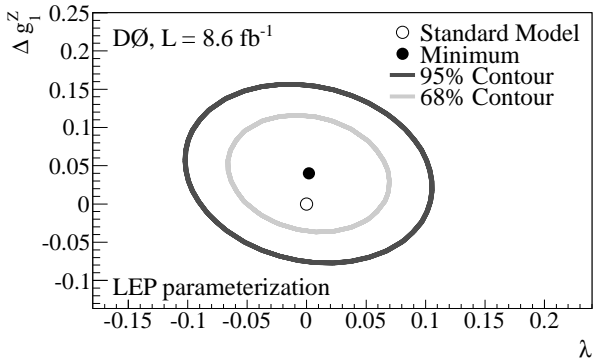}
      \caption{$WZ\rightarrow{\ell\nu}\ell\ell$~($l=\mu,e$). 
	The 68\% and 95\% C.L. two-parameter limits on the $\gamma WW/ZWW$ 
	coupling parameters assuming the LEP parametrization with $\Lambda=2$~TeV.  
	The black circle indicates the most probable values of ATGCs from the 
	two-parameter fit.} 
      \label{fig:limit2} 
    \end{centering} 
  \end{figure}

  The limits are determined from a fit of SM and ATGC contributions 
  to the data using the reconstructed variables: the $p_{T}^{jj}$ 
  distribution from $WW+WZ\rightarrow\ell\nu{jj}$ production, the 
  $p_{T}^{\ell\ell}$ distribution from $WZ\rightarrow\ell\nu{\ell\ell}$ 
  production, the $E_{T}^{\gamma}$ distribution from 
  $W\gamma\rightarrow\ell\nu\gamma$ production, and the $p_{T}$ 
  distributions of the two leptons from $WW\rightarrow\ell\nu\ell\nu$ 
  production.  The $p_{T}^{jj}$ and $p_{T}^{\ell\ell}$ distributions  
  from 4.3~fb$^{-1}$ and 8.6~fb$^{-1}$ analyses, respectively, are 
  shown in Fig.~\ref{fig:kinemWWWZ}.  The $E_{T}^{\gamma}$ and lepton's 
  $p_{T}$ distributions, and the $p_{T}^{jj}$ distribution from 1.1~fb$^{-1}$ 
  analysis can be found elsewhere~\cite{bib:wgamma1,bib:wgamma2,bib:xsec1,bib:lvjj1}.  
  The individual contributions are fit to the data as the in the presence 
  of ATGCs by minimizing the $\chi^2$ function with respect to Gaussian 
  priors on each of the systematic uncertainties~\cite{bib:wadeLim}.  
  The fit is performed simultaneously on kinematic distributions corresponding 
  to the different sub-channels and data epochs.  The remaining $p_{T}^{jj}$ 
  distributions for the electron and muon channels from the 1.1~fb$^{-1}$ 
  $WW+WZ\rightarrow\ell\nu{jj}$ analysis are fit separately and the 
  $\chi^2$ values are summed with those obtained in the simultaneous 
  fit.  The effects of systematic uncertainties on separate samples 
  and sub-channels due to the same uncertainty are assumed to be 100\% 
  correlated but different uncertainties are assumed to be uncorrelated.  
  \begin{table*}[tbp]
    \caption{\label{tab:resultsLVJJ} The 95\%~C.L. one-parameter 
      limits on ATGCs from $WZ\rightarrow\ell\nu\ell\ell$ and $WW+WZ\rightarrow \ell\nu{jj}$ 
      ($l=\mu,e$) final states with $\Lambda=2$~TeV. The analyzed integrated luminosity for 
      each analysis is also presented together with the time period of data collection.}
    \begin{ruledtabular}
      \begin{tabular}{lcccc}
      LEP parametrization & Integrated luminosity & $\Delta\kappa_{\gamma}$ & $\lambda$ & $\Delta{g_{1}^{Z}}$ \\ \hline
      $WZ\rightarrow\ell\nu\ell\ell$     & $8.6~\rm{fb^{-1}}$ (2002~$-$~2011) &        $-$      & $[-0.077, 0.089]$ & $[-0.055, 0.117]$  \\
      $WW+WZ\rightarrow \ell \nu jj$     & $4.3~\rm{fb^{-1}}$ (2006~$-$~2009) & $[-0.27, 0.37]$ & $[-0.075, 0.080]$ & $[-0.071, 0.137]$  \\ 
      \hline\\
      Equal couplings parameterization & Integrated luminosity & $\Delta\kappa$ & $\lambda$ &  \\ \hline
      $WZ\rightarrow \ell\nu \ell\ell$   & $8.6~\rm{fb^{-1}}$ (2002~$-$~2011) &     $-$         & $[-0.077, 0.090]$ & \\ 
      $WW+WZ\rightarrow \ell\nu jj$      & $4.3~\rm{fb^{-1}}$ (2006~$-$~2009) & $[-0.078, 0.153]$ & $[-0.074, 0.079]$ &  \\ 
      \end{tabular}
    \end{ruledtabular}
  \end{table*}
  
  The 68\% and 95\%~C.L. limits on ATGCs from the 4.3~fb$^{-1}$ 
  analysis of $WW+WZ\rightarrow\ell\nu{jj}$ final states in the two-parameter 
  space are shown in Fig.~\ref{fig:limit1}.  The limits from the 8.6~fb$^{-1}$ 
  analysis of $WZ\rightarrow\ell\nu{\ell\ell}$ final states are presented 
  only in the $\lambda -\Delta g_1^Z$ space as shown in Fig.~\ref{fig:limit2}, 
  because $WZ$ production is weakly sensitive to $\Delta\kappa_\gamma$ 
  via the relation given by Eq.~(\ref{eq:lepparam}).  The 95\%~C.L. 
  one-parameter limits, obtained from single parameter fits with all other 
  parameters fixed to their SM values are presented in Table~\ref{tab:resultsLVJJ}.
  
\begin{figure*}[tbp] 
    \begin{centering}
      \includegraphics[width=8.0cm]{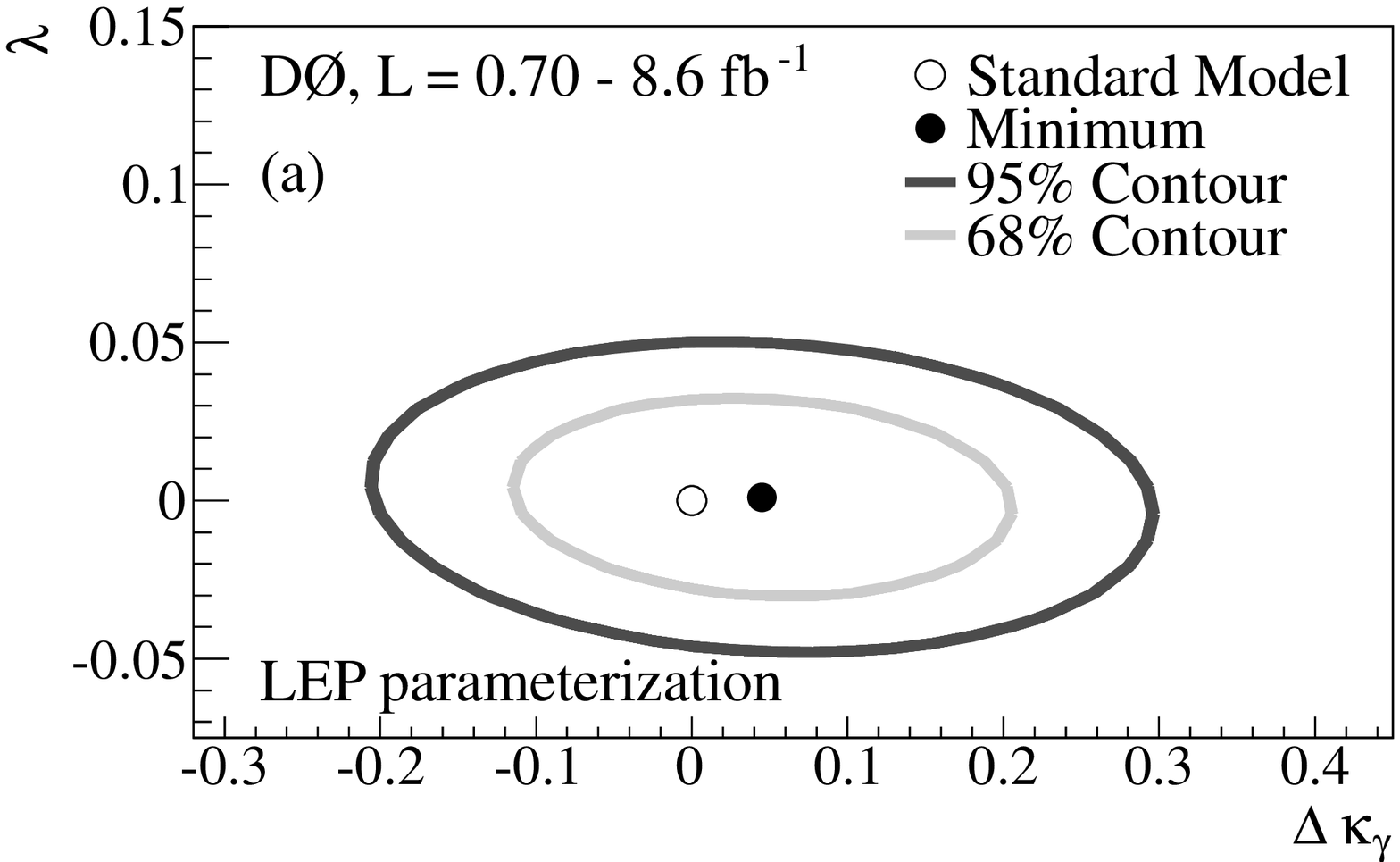}
      \includegraphics[width=8.0cm]{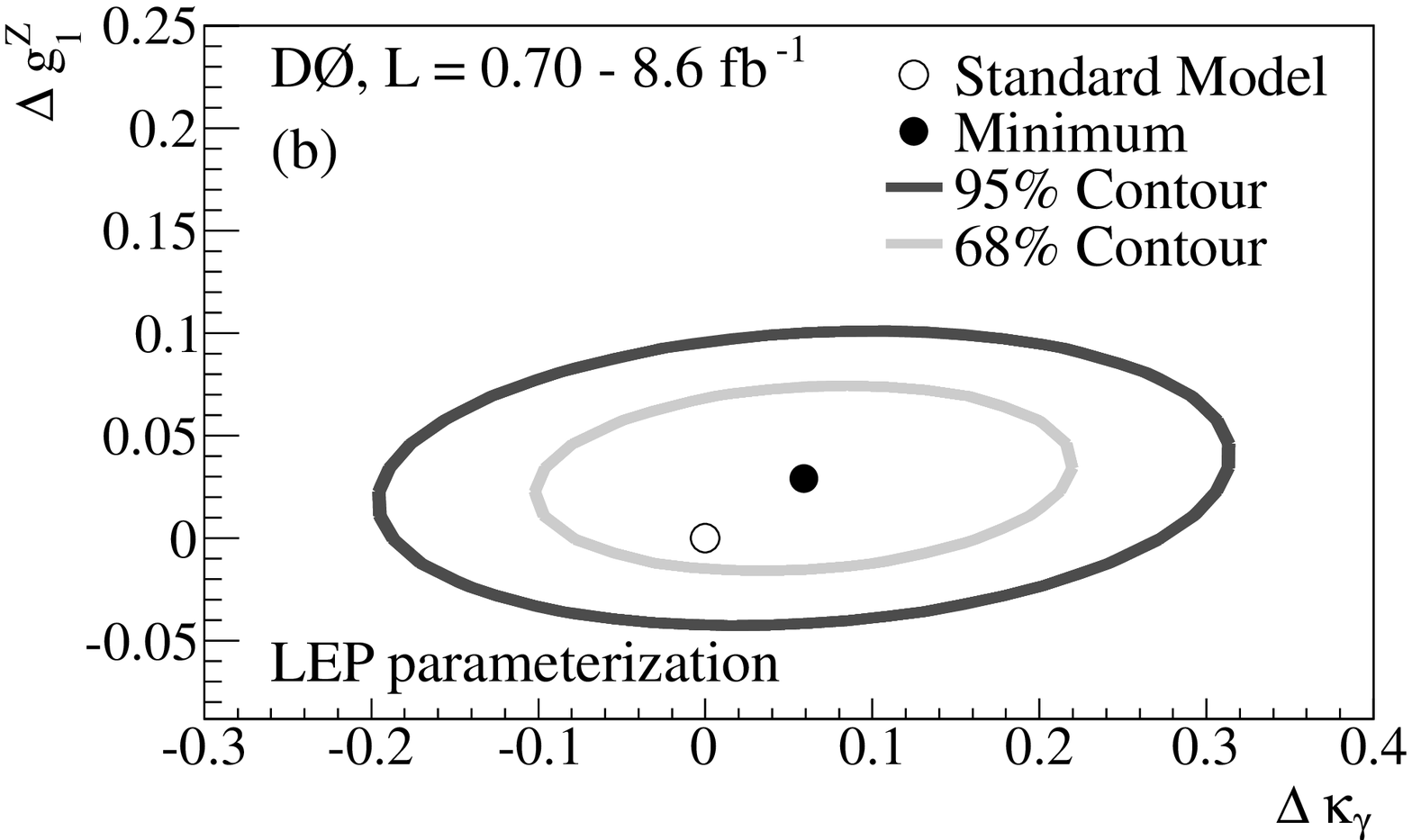}
      \includegraphics[width=8.0cm]{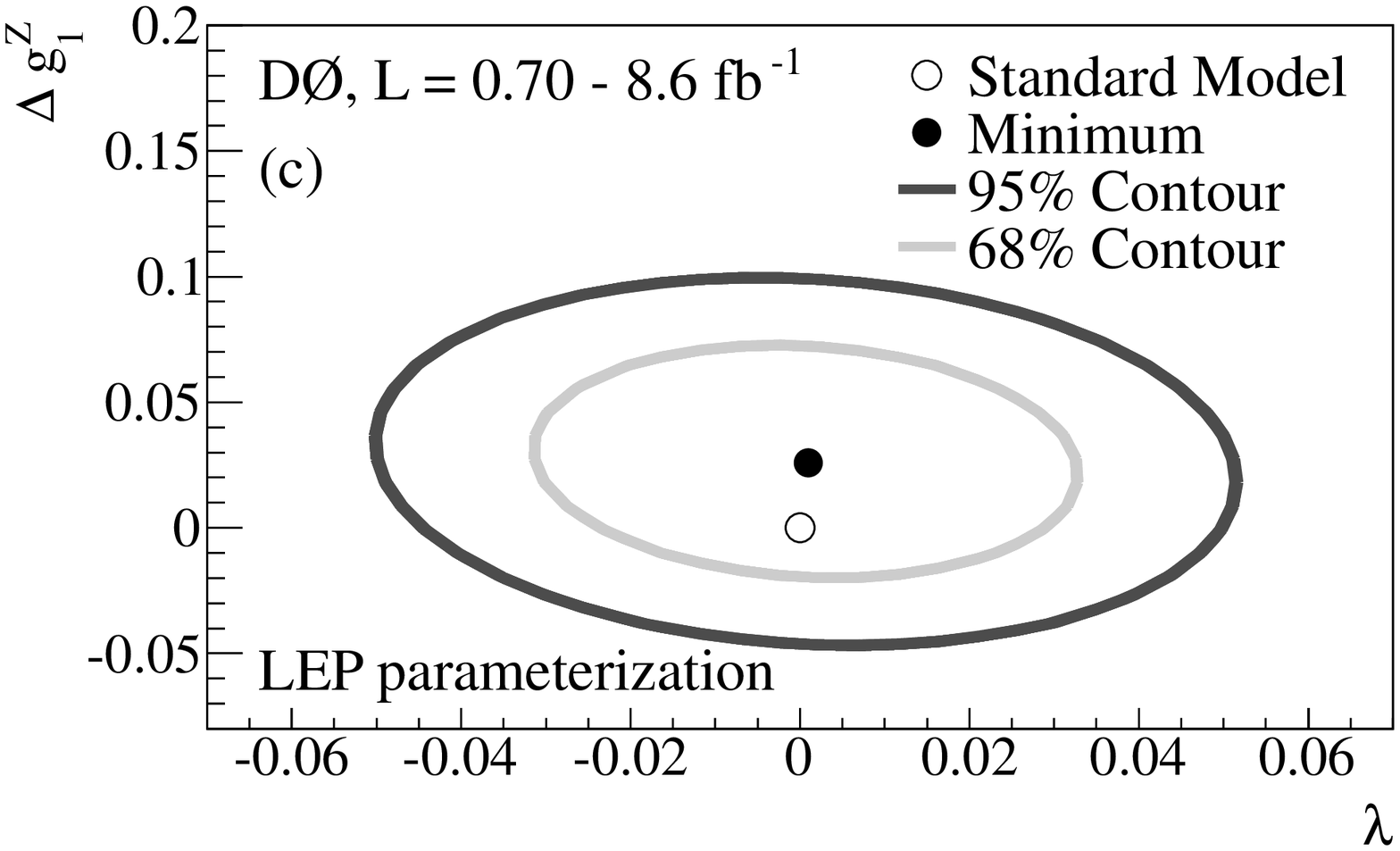}
      \includegraphics[width=8.0cm]{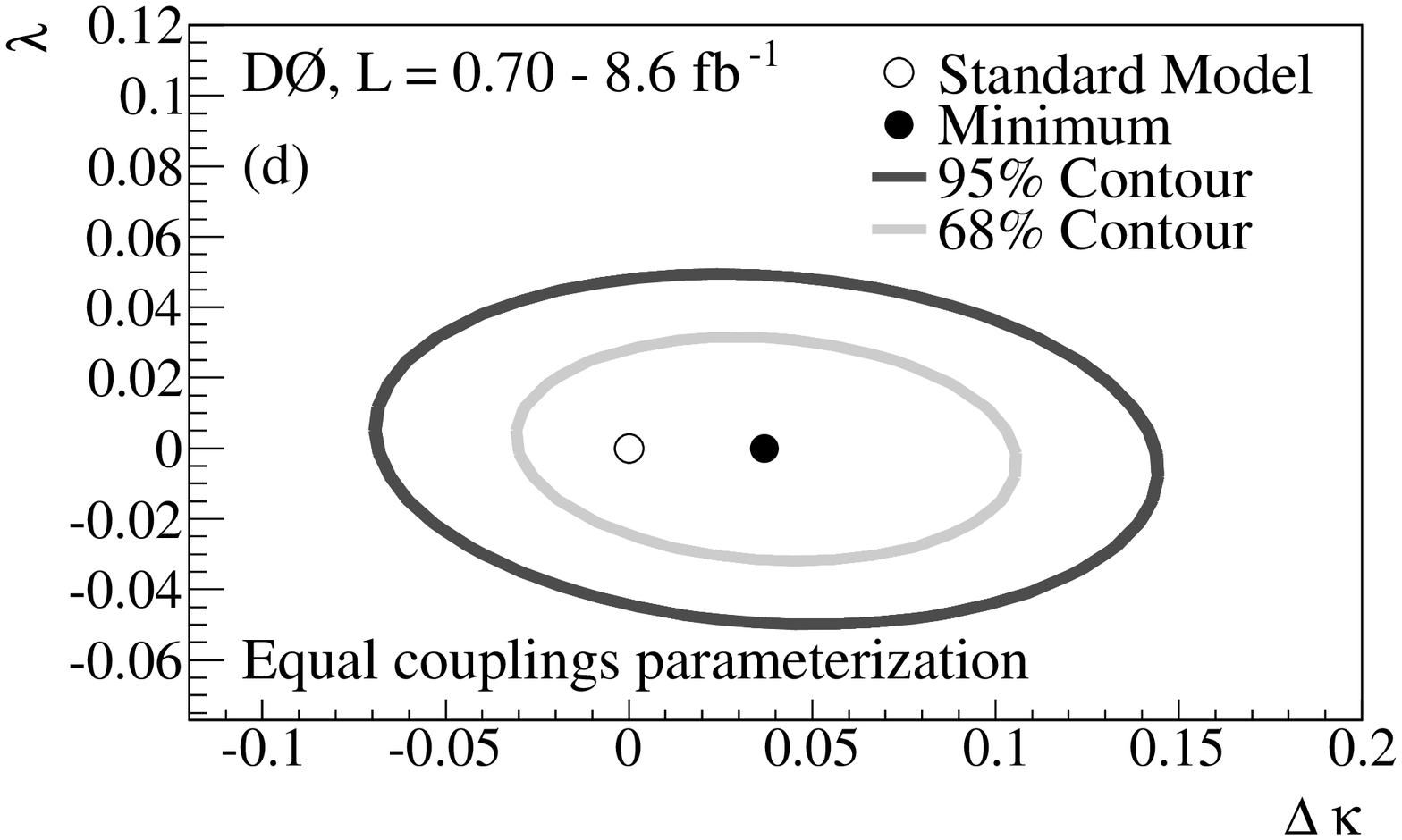}
      \caption{The 68\% and 95\%~C.L. two-parameter limits on the $\gamma WW/ZWW$ 
	ATGCs $\Delta\kappa_{\gamma}$, $\Delta\lambda_{\gamma}$ and $\Delta g_{1}^{Z}$, 
	assuming the LEP parametrization (a,~b,~c) and on $\Delta\kappa$ and $\lambda$ 
	ATGCs for the equal couplings parameterization (d) with $\Lambda=2$~TeV from 
	the combination of $WW+WZ\rightarrow\ell\nu{jj}$, $WZ\rightarrow\ell\nu{\ell\ell}$, 
	$W\gamma\rightarrow\ell\nu\gamma$, and $WW\rightarrow\ell\nu\ell\nu$ final states 
	($l=\mu,e$). Black circles indicate the most probable values of ATGCs from 
        the two-parameter fit.} 
      \label{fig:limit3} 
    \end{centering} 
  \end{figure*}

  The resulting 68\% and 95\%~C.L. one-parameter limits from the 
  combined fit of $\ell\nu\gamma$, $\ell\nu\ell\nu$, $\ell\nu{jj}$, 
  and $\ell\nu{\ell\ell}$ final states are shown in Table~\ref{tab:combo} 
  and limits in two-parameter space are shown in Fig.~\ref{fig:limit3}.  
  The limits in both scenarios represent an improvement relative to 
  previous results from the 
  Tevatron~\cite{bib:run1prd,bib:cdfresult,bib:wgamma1,bib:wgamma2,bib:xsec1,bib:lvjj1}.  
  For the LEP parametrization, our combined measurement with 68\%~C.L. 
  allowed intervals of $\kappa_{\gamma}=1.048^{+0.106}_{-0.105}$, 
  $\lambda=0.007^{+0.021}_{-0.022}$, and $g_{1}^{Z}=1.022^{+0.032}_{-0.030}$ 
  presented in this paper has similar sensitivity to the results from the 
  individual LEP experiments~\cite{bib:aleph,bib:opal,bib:l3,bib:delphi}. 
  The combined D0 limits are more stringent than those set by the ATLAS 
  Collaboration for $\Lambda=2$~TeV~\cite{bib:ATLAS}.  The limits from the 
  CMS Collaboration~\cite{bib:CMS} are not directly comparable to our results 
  due to a different assumption for $\Lambda$ value that affects a dipole 
  form factor and thus, the sensitivity to ATGCs~\cite{bib:bounds}.  
  Nevertheless, the combined D0 limits on $\Delta\kappa_\gamma$, 
  $\lambda$ and $\Delta g_{1}^{Z}$ are more stringent than both ATLAS and 
  CMS current limits for $\Lambda\rightarrow\infty$.
  
  \begin{table}[tbp]
    \caption{\label{tab:combo} One-dimensional $\chi^2$ minimum and 68\% and 95\%~C.L. 
      allowed intervals on anomalous values of $\gamma WW/ZWW$ ATGCs from the combined 
      fit of $WW+WZ\rightarrow\ell\nu{jj}$, $WZ\rightarrow\ell\nu{\ell\ell}$, 
      $W\gamma\rightarrow\ell\nu\gamma$, and $WW\rightarrow\ell\nu\ell\nu$ final 
      states.}
    \begin{ruledtabular}
      \begin{tabular}{lrcc}
	\multicolumn{4}{c}{Results for LEP parameterization} \\
	Parameter & Minimum & 68\%~C.L. & 95\%~C.L. \\ \hline
	$\Delta\kappa_\gamma$	& $0.048$  & $[-0.057, 0.154]$  & $[-0.158, 0.255]$ \\
	$\Delta g_1^Z$		& $0.022$  & $[-0.008, 0.054]$  & $[-0.034, 0.084]$ \\
	$\lambda$	        & $0.007$  & $[-0.015, 0.028]$  & $[-0.036, 0.044]$ \\ \\
	$\mu_W$~$(e/2M_W)$      & $2.012$  & $[1.978, 2.047]$   & $[1.944, 2.080]$ \\
	$q_W$~$(e/M^2_W)$       & $-0.995$ & $[-1.038, -0.953]$ & $[-1.079, -0.916]$ \\
	\hline
	\multicolumn{4}{c}{Results for Equal couplings parameterization} \\
	Parameter & Minimum & 68\%~C.L. & 95\%~C.L. \\ \hline
	$\Delta\kappa$	    & $0.037$  & $[-0.007, 0.081]$  & $[-0.049, 0.124]$ \\
	$\lambda$	    & $0.008$  & $[-0.017, 0.028]$  & $[-0.039, 0.042]$ \\ \\
	$\mu_W$~$(e/2M_W)$  & $2.016$  & $[1.982, 2.050]$   & $[1.948, 2.082]$ \\
	$q_W$~$(e/M^2_W)$   & $-1.009$ & $[-1.050, -0.970]$ & $[-1.092, -0.935]$ \\
      \end{tabular}
    \end{ruledtabular}
  \end{table}
  
  \begin{figure*}[tbp]
    \includegraphics[width=8.0cm]{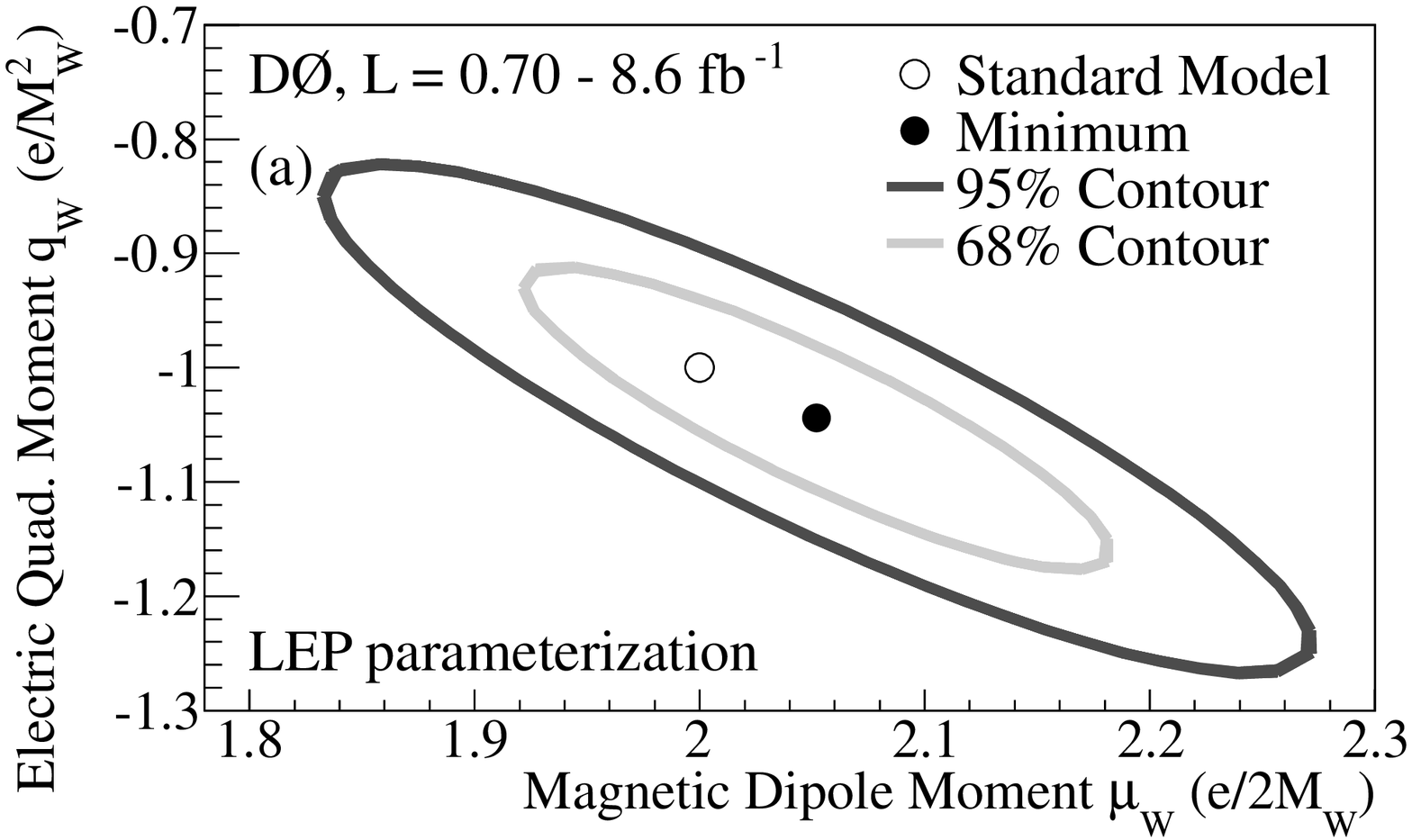}
    \includegraphics[width=8.0cm]{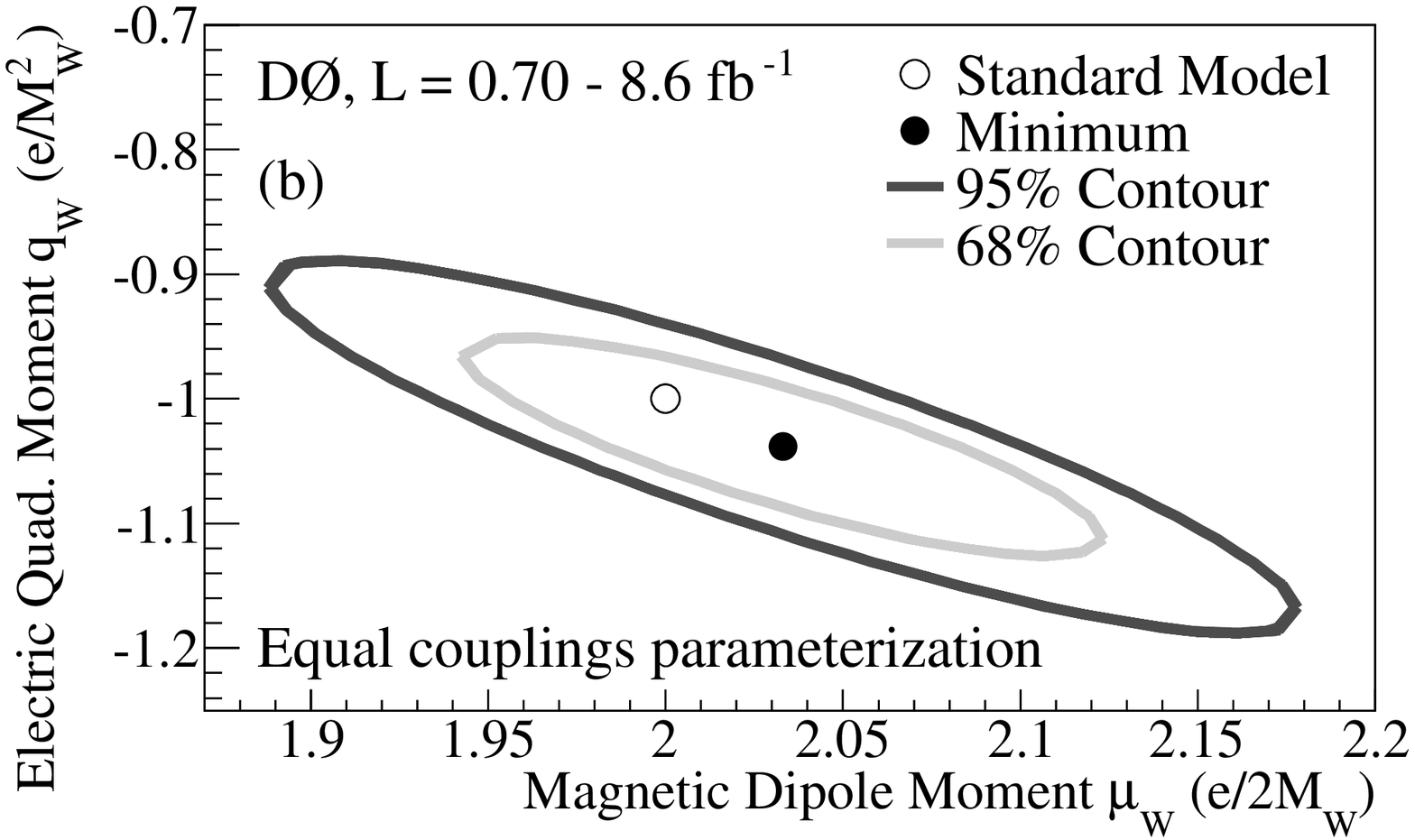}  
    \caption{Two-dimensional 68\% and 95\%~C.L. limits for the $W$ boson electric quadrupole 
      moment vs.\ the magnetic dipole moment for (a) LEP parametrization and (b) equal 
      couplings constraints from the combination of $WW+WZ\rightarrow\ell\nu{jj}$, 
      $WZ\rightarrow\ell\nu{\ell\ell}$, $W\gamma\rightarrow\ell\nu\gamma$, and 
      $WW\rightarrow\ell\nu\ell\nu$ final states ($l=\mu,e$).  In both cases we 
      assume $\Lambda=2$~TeV. Black circles indicate the most probable values of $\mu_W$ 
      and $q_W$ from the two-parameter fit.}
    \label{fig:2dmom}
  \end{figure*}
  
  \noindent Using observed limits we extract measurements of the $W$ boson 
  magnetic dipole and electric quadrupole moments.  When assuming 
  the LEP parameterization with $g^Z_1=1$, we set the 68\%~C.L. 
  intervals of $\mu_W=2.012^{+0.035}_{-0.034} \, (e/2M_W)$ 
  and $q_W=-0.995^{+0.042}_{-0.043} \, (e/M^2_W)$.  The 68\% 
  and 95\%~C.L. limits on $\mu_W$ and $q_W$ in both scenarios are shown 
  in Fig.~\ref{fig:2dmom}.  

  In summary, we have presented new searches of anomalous 
  $\gamma{WW}$ and $ZWW$ trilinear gauge boson couplings from 
  $WW+WZ\rightarrow\ell\nu{jj}$ and $WZ\rightarrow\ell\nu{\ell\ell}$ 
  channels analyzing 4.3~fb$^{-1}$ and 8.6~fb$^{-1}$ of integrated 
  luminosity, respectively, and we set limits on ATGCs for 
  these final states.  The limits from 4.3~fb$^{-1}$ $\ell\nu{jj}$ 
  analysis are the best	limits to date at a hadron collider in 
  this final state.  The limits from 8.6~fb$^{-1}$ $\ell\nu\ell\ell$ 
  analysis are comparable to those set at the LHC and improve 
  relative to previous limits set in this final state at the Tevatron~\cite{bib:wzTeV}.  
  We have combined these results with those previously published from 
  $WW+WZ\rightarrow\ell\nu{jj}$~(1.1~fb$^{-1}$), 
  $W\gamma\rightarrow\ell\nu\gamma$~(4.9~fb$^{-1}$), and 
  $WW\rightarrow\ell\nu\ell\nu$~(1.0~fb$^{-1}$) final states using 
  up to 8.6~fb$^{-1}$ of integrated luminosity.  No deviation from 
  the SM is found in data.  We set the most stringent limits on 
  $\Delta\kappa_\gamma$, $\lambda$ and $\Delta g_{1}^{Z}$ at a 
  hadron collider to date complementing similar measurements 
  performed at LEP and LHC.  Using the LEP parameterization we set 
  the combined 68\%~C.L. limits of 
  $-0.057<\Delta\kappa_\gamma<0.154$, $-0.015<\lambda<0.028$, 
  and $-0.008<\Delta g_1^Z<0.054$. At 95\%~C.L. the limits are 
  $-0.158<\Delta\kappa_\gamma<0.255$, $-0.036<\lambda<0.044$, 
  and $-0.034<\Delta g_1^Z<0.084$.  Based on the combination of 
  all diboson production and decay channels we set the most 
  stringent 68\%~C.L. constraints on the $W$ boson magnetic dipole 
  and electric quadrupole moments of 
  $\mu_W=2.012^{+0.035}_{-0.034} \, (e/2M_W)$ and 
  $q_W=-0.995^{+0.042}_{-0.043} \, (e/M^2_W)$, respectively, to date.

%
We thank the staffs at Fermilab and collaborating institutions,
and acknowledge support from the
DOE and NSF (USA);
CEA and CNRS/IN2P3 (France);
MON, NRC KI and RFBR (Russia);
CNPq, FAPERJ, FAPESP and FUNDUNESP (Brazil);
DAE and DST (India);
Colciencias (Colombia);
CONACyT (Mexico);
NRF (Korea);
FOM (The Netherlands);
STFC and the Royal Society (United Kingdom);
MSMT and GACR (Czech Republic);
BMBF and DFG (Germany);
SFI (Ireland);
The Swedish Research Council (Sweden);
and
CAS and CNSF (China).


\begin{thebibliography}{99}
\bibitem{bib:newref} Limits on anomalous couplings presented in this paper are given as the low energy limits of the couplings.
\bibitem{bib:def} D0 uses a coordinate system with the $z$ axis running along the proton beam axis. The angles $\theta$ and $\phi$ are the polar and azimuthal angles, respectively. Pseudorapidity is defined as $\eta = -\ln \left[\tan(\theta/2) \right]$, where $\theta$ is measured with respect to the proton beam direction.
\end{thebibliography}
 \end{document}